%% file: main.tex
\documentclass{egpubl}
\usepackage{booktabs} % For formal tables
\usepackage{textcomp}
\usepackage{amsmath}
\usepackage{amsfonts}
\usepackage{duckuments}
\usepackage{bm}
\usepackage{url}

\DeclareMathOperator*{\argmin}{arg\,min}

\JournalSubmission    % uncomment for submission to Computer Graphics Forum
\usepackage[T1]{fontenc}
\usepackage{dfadobe}  

\usepackage{cite}  % comment out for biblatex with backend=biber
\BibtexOrBiblatex
\electronicVersion
\PrintedOrElectronic
\ifpdf \usepackage[pdftex]{graphicx} \pdfcompresslevel=9
\else \usepackage[dvips]{graphicx} \fi

\usepackage{egweblnk} 
\title[View-Dependent Formulation of 2.5D Cartoon Models]%
      {View-Dependent Formulation of 2.5D Cartoon Models}

\author[T. Fukusato \& A. Maejima]
{
    \parbox{\textwidth}{\centering T.\,Fukusato$^{1}$
    and A. Maejima$^{2, 3}$} \\
    {\parbox{\textwidth}{\centering $^1$The University of Tokyo \hspace{2mm}
    $^2$OLM Digital Inc.
    \hspace{2mm}
    $^3$IMAGICA GROUP Inc.}
    }
}
\begin{document}
% uncomment for using teaser
% \teaser{
%  \includegraphics[width=\linewidth]{eg_new}
%  \centering
%   \caption{New EG Logo}
% \label{fig:teaser}
%}

%\teaser{
%  \centering
%\includegraphics[width=0.98\linewidth]{fig/teasor_v2.pdf}
%\caption{The concept of our system. We envision a two-and-a-half-dimensional (2.5D) graphics operator for generating 3D-like effects in an arbitrary viewpoint based on a set of 2D shapes from several viewpoints, such as front and side views, inspired by view-dependent deformation. Note that the input cartoon character was designed using Harmony~\protect\shortcite{toonboom2020}.\vspace{3mm}} 
%\label{fig:teasor} 
%}

\maketitle
%-------------------------------------------------------------------------
\begin{abstract}
2.5D cartoon models~\cite{rivers20102, kitamura2014modeling, gois2015interactive, coutinho2016puppeteering} are methods to simulate three-dimensional (3D)-like movements, such as out-of-plane rotation, from two-dimensional ($2$D) shapes in different views. 
However, cartoon objects and characters have several distorted parts which do not correspond to any real 3D positions (e.g., Mickey Mouse's ears), that implies that existing systems are not suitable for designing such representations. 
Hence, we formulate it as a view-dependent deformation (VDD) problem~\cite{chaudhuri2004system, chaudhuri2007reusing, koyama2013view, rademacher1999view}, which has been proposed in the field of 3D character animation. 
The distortions in an arbitrary viewpoint are automatically obtained by blending the user-specified 2D shapes of key views. This model is simple enough to easily implement in an existing animation system. Several examples demonstrate 
the robustness of our method over previous methods. In addition, we conduct a user study and confirm that the proposed system is effective for animating classic cartoon characters.


\end{abstract}  

%-------------------------------------------------------------------------
\input{sec/1_introduction.tex}
\input{sec/2_relatedwork.tex}

\input{sec/3_ui.tex}
\input{sec/4_method.tex}
\input{sec/5_result.tex}
\input{sec/6_userstudy}

\input{sec/7_limitation}

\input{sec/8_conclusions}

%-------------------------------------------------------------------------
% bibtex
\bibliographystyle{eg-alpha-doi} 
\bibliography{main.bib}       

\end{document}

%% file: sec/1_introduction.tex
\section{Introduction}
\label{sec:intro}
Computer-assisted techniques known as three-dimensional ($3$D) character animation are widely used for improving the efficiency of rotoscoping in the classic $2$D cartoon. However, the cartoon drawings are not bound to geometric precision and typically contain many subtle artistic distortions, such as changes in scale and perspective, or more noticeable effects, such as changes in the shape or location of features (for example, Mickey Mouse's ears always face forward from any view, and the one ear changes positions and is shifted downward). 
When creating such animation with 3D, these gaps would be filled by additional deformation by artists in each view. 
Therefore, instead of conventional 3D models, our goal is to establish a simple framework to make 3D-like movements specialized for 2D cartoons. %That is, 
%Therefore, instead of conventional 3D models, our goal is to establish an artist-friendly framework to make 3D-like movements specialized for 2D cartoons. %That is, 

Some pioneering work show simulating 3D-like rotations from 2D drawings in anterior and lateral views~\cite{di2001automatic, Furusawa2014QR, live2019, reallusion2020cartoon, yeh2012double}. They utilize a simple heuristic approximating an object as a sphere. However, their method are not robust, nor are easy to represent to cartoon-like rotation (with artistic distortions). Rivers et al.~\shortcite{rivers20102} propose a novel hybrid structure of 2.5D graphics (i.e., 2D plus depth information).
This structure associates each part of a 2D drawing with a single 3D anchor position, and it can estimate the 2D part's position and the depth value in a new view. Note that the shape of each part is interpolated in 2D. Although the concept is effective for generating 3D-like movements in new views, if several parts do not correspond to any real 3D position (e.g., Mickey Mouse's ears), the user must manually change the parts' positions based on 2D-space interpolation without 3D anchors. Rivers et al. does not argue how to handle the artistic distortions in multi-view 2D drawings.  

Following the spirit of the recent work in enriching cartoon animation, we formulated 2.5D graphics with artistic distortions of multi-view 2D drawings and developed a user interface to manually design 2.5D cartoon models from scratch. In summary, our work contains the following key contributions:

\noindent
\begin{itemize}
%\setlength{\leftskip}{-3mm}
%\item A user interface for simulating 2.5D graphics based on a set of 2D shapes from various viewpoints.
\item A novel method to estimate anchor positions, shapes, and depth values of parts in an arbitrary view with a view-dependent deformation (VDD)-based mechanism. %control
\item Professional and amateur artists' feedback demonstrating the benefit of our 2.5D cartoon models.
\end{itemize}

\begin{figure*}[t]
\centering
\includegraphics[width=0.9\linewidth]{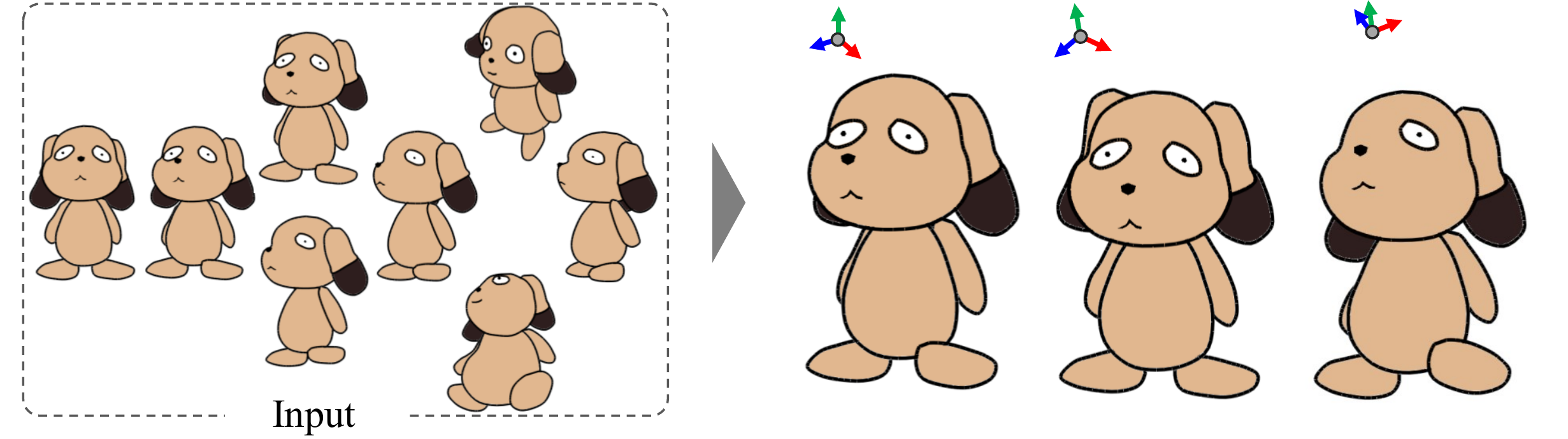}
\caption{The concept of our system. We envision a two-and-a-half-dimensional (2.5D) graphics operator for generating 3D-like effects in an arbitrary viewpoint based on a set of 2D shapes from several viewpoints, such as front and side views, inspired by view-dependent deformation. Note that the input cartoon character was designed using Harmony~\protect\shortcite{toonboom2020}.} 
\label{fig:teasor} 
\end{figure*}
%\vspace{3mm}

%% file: sec/2_relatedwork.tex
\section{Relatedwork}
\label{sec:relatedWork}
This section reviews prior work on frameworks for constructing conventional 2D/3D graphics and designing cartoon-like 2.5D graphics from 2D/3D models.

\subsection{2D/3D Graphics System}
For animating 2D drawings, existing approaches typically take one of two approaches. One approach is to deform input 2D images by using simple deformers~\cite{hornung2007character, igarashi2005rigid, su2018live, xing2015autocomplete} or physically-inspired simulation~\cite{kazi2016motion, willett2017secondary, xing2016energy}, and switching them like a flip comic~\cite{furukawa2017voice, morimoto2019generating, willett2017triggering, willett2020pose}.
The other approach is to blend the interiors of the given 2D shapes, such as two-shape interpolation~\cite{alexa2000rigid, baxter2008rigid, baxter2009compatible, chen2013planar, kaji2012mathematical, yang2019cr}, examples-way blending~\cite{baxter2006latent, baxter2009n, bregler2002turning, Igarashi2005SKP, ma2012blendshape}, and stroke-based interpolation~\cite{de2006re, fukusato2016active, sederberg19932, sederberg1992physically, whited2010betweenit, yang2017context, yang2018ftp}. 
Their morph is as-rigid-as-possible (ARAP) in the sense that local volumes (areas) are least-distorting as they vary from their source to target configurations. Nonetheless, these algorithms focus only on 2D-space transformation, so 3D-like movements, such as 3D rotations, are still difficult to control.

In 3D graphics, various approaches have been proposed for generating $3$D-specific models, such as human faces~\cite{blanz1999mms, han2017deepsketch2face, reallusion2019crazy}, rounded objects~\cite{dvorovzvnak2020monster, gingold2009structured, igarashi1999teddy}, humanoid characters~\cite{bessmeltsev2015MCC, levi2013artisketch}, or constructive solid geometry models~\cite{rivers2010modeling} from $2$D images (or sketches). However, the resulting 3D models do not retain the detail of $2$D drawings, such as artistic effects, at all, and thus, they are not suitable for designing $2$D cartoons. 

%--------------------------------------------------------------------------
\subsection{2.5D Graphics System}
\label{related:2.5D}

\subsubsection{2D to 2.5D Graphics}
Several systems enable the user to handle self-occlusion and generate new $3$D-like views and movements from $2$D drawings ($2$D $\rightarrow$ $2.5$D) by splitting the drawing parts into layers (i.e., vector graphics representation) and deforming them~\cite{carvalho2017dilight, boris2015vector, di2001automatic, Furusawa2014QR, live2019, reallusion2020cartoon, yeh2012double}. %
However, these systems accept a very limited variation of character shape (i.e., sphere-like model) and pose. The underlying cause is that the designing process of 3D-like rotation of 2D cartoons remains a highly time-consuming task that requires artistic know-how (e.g., redrawing some or all of the models); hence, their sphere-based calculation is not suitable. 
Rivers et al.~\shortcite{rivers20102} propose a hybrid structure that consists of 2D drawings (vector graphics), each associated with a 3D-space anchor. Given multi-view 2D drawings, they do not approximate a 3D model but rather estimate the 3D anchor position of each part. The output shape is obtained by the 2D-space deformation system. These kinds of 2.5D graphics have attracted attention as a new approach to cartoon control, and several extensions have been proposed~\cite{coutinho2016puppeteering, kitamura2014modeling, gois2015interactive}. 
However, their research papers mainly focus on the concept of 2.5D graphics, and the view-specific distortions in classical $2$D cartoons are not discussed much. In addition, their system requires much time-consuming manual work.

We therefore build on their concept but provide a description of the algorithm that is clearer and easier to formulate while handling the view-specific distortions. 

%--------------------------------------------------------------------------
\subsubsection{3D to 2.5D Graphics}
On the topic of using 3D models, stylized shading techniques are often used to emulate the style of classic 2D cartoons~\cite{barla2006x, sloan2001lit, todo2007locally} (3D $\rightarrow$ 2.5D). These systems assume that the shading effects are described by view-space normals and can produce various scenes beyond traditional 3D lighting control. 

In the context of 3D geometry-processing research, VDD is a popular way of discussing the artistic distortions of classical 2D cartoons~\cite{rademacher1999view}. For a given set of reference deformations of key viewpoints, this system deforms a 3D base model in a new viewpoint by automatically blending the deformations. In addition, the VDD mechanism is also extended to animation systems such as keyframe animation~\cite{chaudhuri2004system, chaudhuri2007reusing} and position-based dynamics~\cite{koyama2013view}. 
However, one problem with these techniques is that the user must prepare in advance a 3D base model and several 3D examples that are associated with a particular view direction. This requires manual intervention and can be especially difficult when creating the 3D cartoon models and the reference deformations.

We therefore extend these mechanisms to apply a 2D layered model and handle the view-specific distortions.

%% file: sec/3_ui.tex
%--------------------------------------------------------------------------
\section{User Interface} %System Overview
\begin{figure}[t]
\centering
  \includegraphics[width=1.0\linewidth]{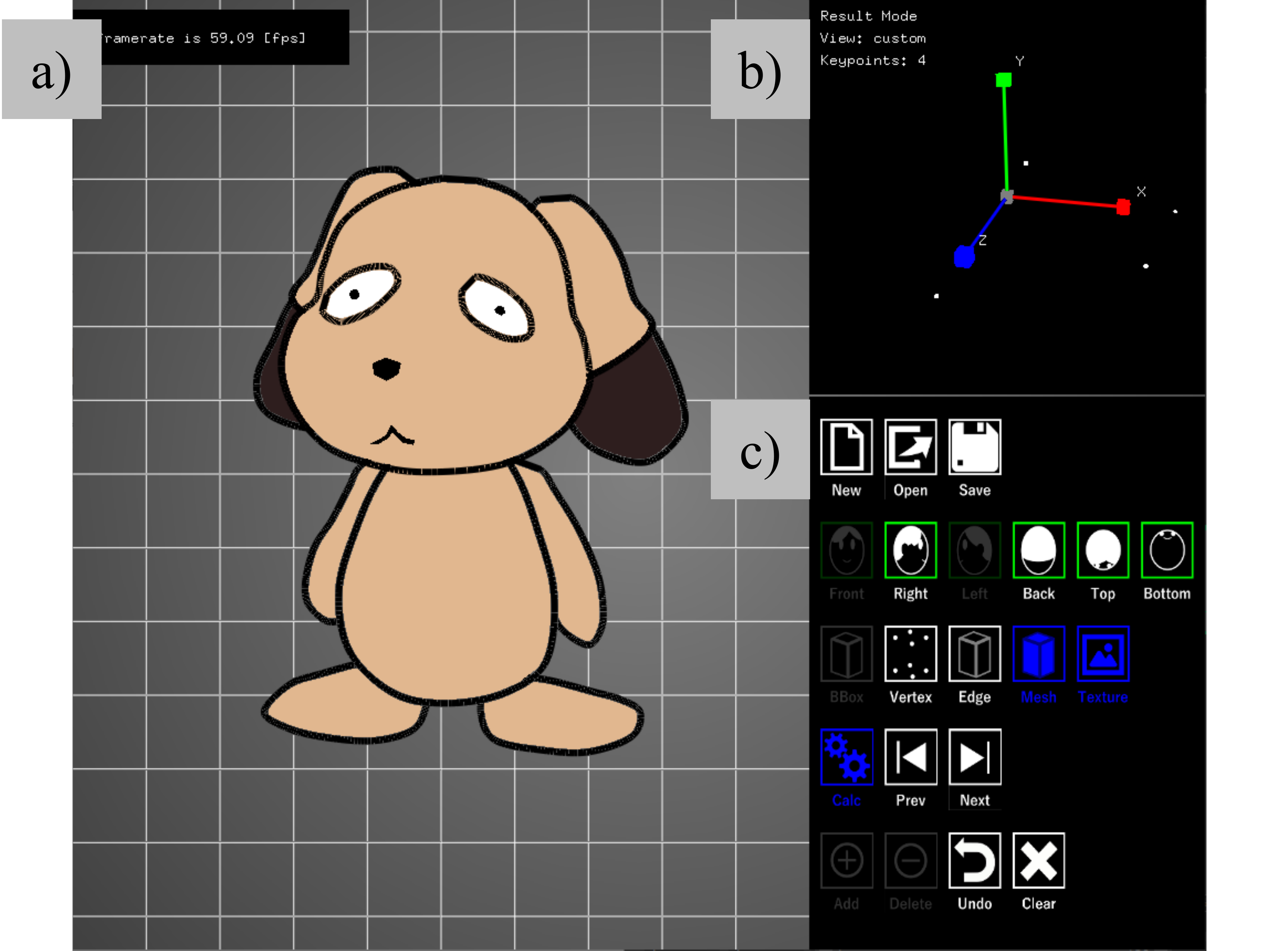}
  \caption{Screenshot of the proposed system that identifies (a) the modeling panel, (b) the view control panel, and (c) tool buttons.}
%\vspace{-2mm} 
\label{fig:screenshot}
\end{figure}

\label{sec:ui}
In this section, we describe how users interact with our system (see Figure~\ref{fig:screenshot}) to design a 2.5D model from multi-view 2D shapes. The user interface of our prototype system shares the similar visual design to previous 2.5D modeling systems~\cite{coutinho2016puppeteering, gois2015interactive, rivers20102} and standard modeling systems~\cite{live2019, reallusion2020cartoon}. First, we begin the process of manually preparing multi-view 2D shapes -- already triangulated compatibly -- from various viewpoints. 

%In this section, we describe how users interact with the proposed system (see Figure~\ref{fig:screenshot}) to design a $2.5$D model from multi-view 2D shapes. 
%First, we begin the process of manually preparing multi-view 2D shapes -- already triangulated compatibly -- from various viewpoints.  

%--------------------------------------------------------------------------
%--------------------------------------------------------------------------

%--------------------------------------------------------------------------
\subsection{2D-Space Management}
\label{sec:management}
The user loads a single-view 2D layered model (which is already split into several parts), and the system automatically computes a center position of each part as a 2D anchor position. Second, as with the commercial modeling software~\cite{maya2019, blender2019}, the system enables the user to adjust the shape of each part with simple deformers (e.g., translation, rotation, uniform scaling, and vertex editing).
Note that the user can also make mesh models from image contour lines (e.g., 2D drawings using Harmony~\shortcite{toonboom2020}) based on polygon triangulation methods~\cite{baxter2009compatible}.

%--------------------------------------------------------------------------
\subsection{Viewpoint Selection}
\label{sec:viewSelection}
\begin{figure}[ht]
\centering
  \includegraphics[width=0.8\linewidth]{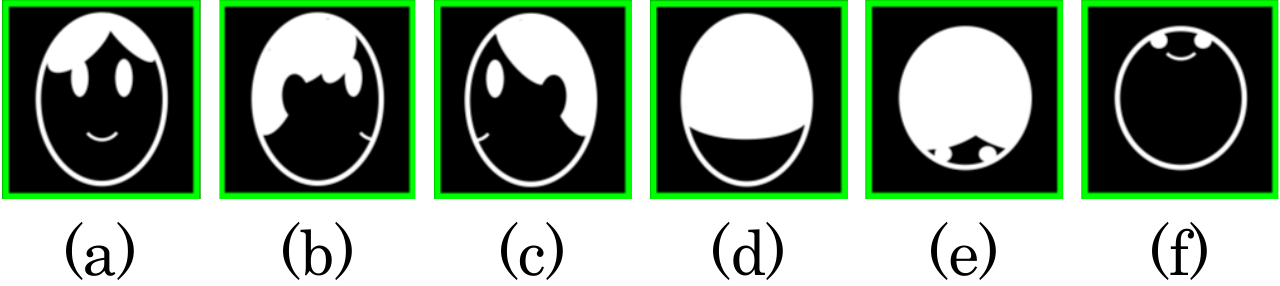}
  \caption{The predefined view buttons, which consist of (a) front, (b) right, (c) left, (d) back, (e) top, and (f) bottom views.}
%\vspace{-2mm} 
\label{fig:view}
\end{figure}

The user clicks an icon to specify the view (see Figure~\ref{fig:view}), and the proposed system associates the layered model to a view direction (front, right, left, back, top, or bottom view). Optionally, the user can choose an arbitrary viewpoint by performing a mouse-drag operation on the view control panel.

%--------------------------------------------------------------------------

%-------------------------------------------------------------------------
\subsection{Key Viewpoint Setting}
\label{sec:key-viewpoint}
After managing the shapes and selecting viewpoints, the user can record these key viewpoint data by clicking the \textit{Add} button. The system shows the key viewpoints (white dots) on the view control panel. In contrast, the \textit{Delete} function deletes the latest key viewpoint data. By repeating the above process, the user can generate multi-view 2D shapes. 

%--------------------------------------------------------------------------
\subsection{Generation and Visualization Function}
\label{sec:generation}
When the user clicks on the \textit{Calc} button, the system automatically generates a 2.5D cartoon model from a set of key viewpoint data and displays it on the modeling panel. 
After generating 2.5D cartoon models, the user can freely turn the generated model around by performing the mouse-drag operation on the view control panel. In addition, by clicking the viewpoint buttons (see Figure~\ref{fig:view}) or the key viewpoints on the view control panel (white dots), the system instantly jumps to the selected view results.

The generation process is sufficiently fast to return immediate feedback (in less than a second) when the user edits key viewpoint data, including additions and deletions. 

%% file: sec/4_method.tex
\section{Methods}
\label{sec:method}
\subsection{2.5D Cartoon Model Generation}
We separated the 2D parts in the key views into two components: (i)~the 3D-space anchor and (ii)~the view-specific distortions.

%--------------------------------------------------------------------------
\subsubsection{Estimating 3D-Space Anchor Positions}
\label{sec:method1}
As with Rivers's system~\cite{rivers20102}, given a set of key viewpoint data, 3D anchor positions of $i$-th parts $\bm{v}^{i} \in \mathbb{R}^{3}$ can be computed using the triangulation algorithm from orthographic projection $\Pi : \mathbb{R}^{3} \rightarrow \mathbb{R}^{2}$ as follows:
\vspace{2mm}
\begin{equation}
%\argmin_{n}
\argmin_{\bm{v}^{i}} \sum_{j \in K} \: \|\bm{v}^{i}_{j} - \Pi \: ( R_{j} \bm{v}^{i})\|^2 
\label{eq:1}
\end{equation}

\noindent
where $K$ is the number of key viewpoint data, $R_{j}$ is the rotation matrix of the $j$-th viewpoint, and $\bm{v}_{j}^{i} \in \mathbb{R}^{2}$ is the 2D anchor position of the $i$-th part in the $j$-th viewpoint. 
Note that if the user sets only the single key viewpoint data (when $|K| = 1$), our system sets depth values of the anchors in its key view at zero. 
%Note that if the user sets only the single key viewpoint data (when $|K| = 1$), our system sets depth values of the anchors in its key views at zero.
%Note that if the user sets only the single key viewpoint data (when $|K| = 1$), our system assigns a depth value to the anchor in its key views using the inflated method~\cite{igarashi1999teddy}. %instead of the above estimation. 

%--------------------------------------------------------------------------
\subsubsection{Estimating 2D-Space Anchor Distortions} % in Key Views
\label{sec:method2}
The projected positions of the estimated 3D anchors on each key viewpoint might be displaced from the 2D anchor positions of the input 2D shapes (when $|K| \geq 2$). 
Then, we focus on a difference in the $j$-th viewpoint $\bm{d}^{i}_{j} \in \mathbb{R}^{2}$ and define it as a 2D-space anchor distortion. %in the key viewpoint.
%We define that the difference in the $j$-th viewpoint $\bm{d}^{i}_{j} \in \mathbb{R}^{2}$ is a 2D-space anchor distortion in the key viewpoint.
%
\vspace{2mm}
\begin{equation}
\bm{d}^{i}_{j} = \bm{v}^{i}_{j} - \Pi \: ( R_{j} \bm{v}^{i}) 
%\bm{d}_{i}^{j} = \bm{v}_{i}^{j} - \Pi \: ( R^{j} \bm{v}_{i}) 
\label{eq:2}
\end{equation}

\begin{figure}[ht]
\centering
  \includegraphics[width=1.0\linewidth]{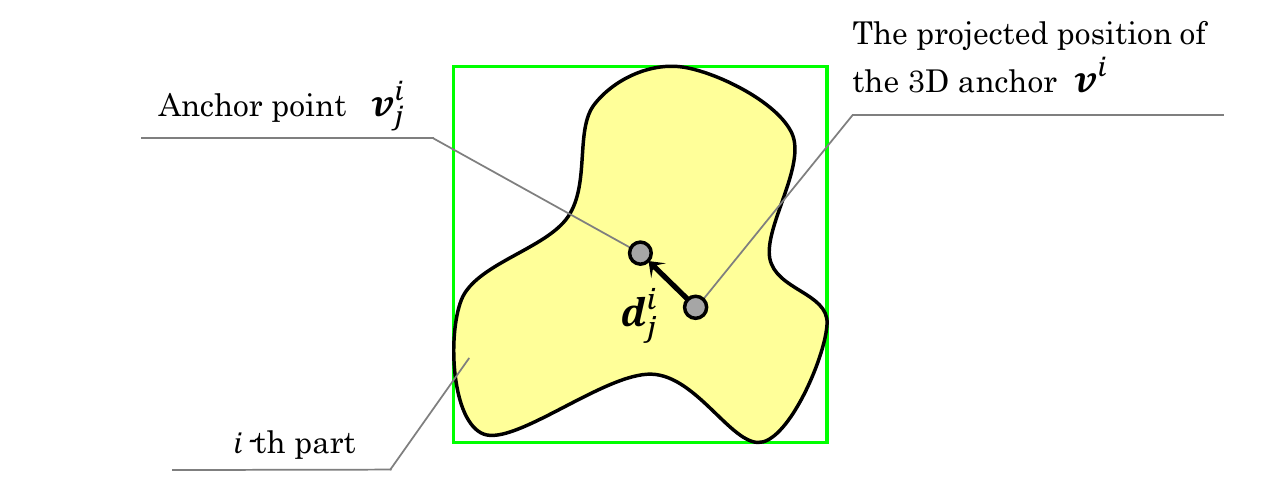}
  \caption{An example of a view-space distortion in key view. The system constructs a set of 2D distortions $\bm{d}_{j}^{i}$ between the user-managed anchor position~$\bm{d}_{j}^{i}$ and the projected position of the estimated 3D anchor~$\bm{v}^{i}$.}
\label{fig:distortion}
\end{figure}

\noindent
Figure~\ref{fig:distortion} illustrates a view-space distortion of anchor in the $j$-th key viewpoint. We denote that when $|K| = 1$, the distortions of 2D parts in all of the views are set to zero~($\|\bm{d}_{j}^{i}\| = 0.0$).
%\leq

%--------------------------------------------------------------------------
\subsection{View-Dependent Control}
\label{sec:VDD}
Given an arbitrary viewpoint $R_{cur}$ in the view control panel, we computed the anchor positions (including the depth values) and the shapes in $2.5$D space by blending a set of the viewpoint data. Therefore, the user must specify appropriate blending weights $w_{j} \in [0.0, 1.0]$, $\sum_{j} w_{j} = 1.0$. Rivers et al.~\shortcite{rivers20102} parameterize the view angles into a 2D rectangular grid (yaw and pitch) and find the $k$-nearest key views on the space. However, their discretization does not consider the roll value. In addition, the previous VDD methods~\cite{chaudhuri2004system, chaudhuri2007reusing, koyama2013view, rademacher1999view} utilize the camera positions in the key viewpoints (3D space), but these remain too difficult to consider rotational information of the key views. Therefore, we employed the camera's rotational matrices in the view control panel and compute the weights based on the Frobenius distance between two rotational matrices of the current view $R^{cur}$ and the $j$-th key view $R^{j}$ as follows:
\begin{equation}
{w}_{j} = \frac{\phi_{j}}{\sum_{k \in K} \phi_{k}} 
\label{eq:3}
\end{equation}
\begin{displaymath}
\phi_{j} = \|R_{cur} - R_{j}\|_{F}^{\alpha}
\end{displaymath}
%\vspace{2mm}

\noindent
where $\| \cdot \|_{F}$ is the Frobenium norm, and $\alpha$ is the constant value (where $\alpha = -4$).

%--------------------------------------------------------------------------
\subsubsection{2.5D-Space Anchor Interpolation}
\label{sec:VDD1}
For estimating the anchor positions of $i$-th parts in a new view, we blended the anchor distortions in the key viewpoints and sum up the projected result of the 3D anchor as follows:
\vspace{2mm}
\begin{equation}
\bm{v}_{cur}^{i} = \Pi \:\: [ R_{cur} \bm{v}^{i} + \sum_{j \in K} w_{j} \: (R_{cur} \cdot R^{-1}_{j})\:  \bm{d}_{j}^{i}] 
\label{eq:4}
\end{equation}
\noindent
The depth value of each part $l_{i}$ is given by 
\vspace{2mm}
\begin{equation}
l^{i} = (R_{cur} \bm{v}^{i})_{z}
\end{equation}
%\vspace{2mm}
%
\noindent
With the above method, the system can assign all the depth values to layer parts without labor-intensive manual depth editing in 3D. Of course, if the user sets the depth values of 2D parts in key views, the system can similarly blend them.

%--------------------------------------------------------------------------
\subsubsection{2.5D-Space Shape Interpolation}
\label{sec:VDD2}
The 2D shape of each layered part $P = (\bm{p}_{0}, \bm{p}_{1}, \cdots, \bm{p}_{n})$, $\bm{p}_{i} \in \mathbb{R}^{2}$ in the current view is determined by ARAP-based interpolation techniques~\cite{alexa2000rigid, baxter2008rigid, baxter2009compatible, chen2013planar, kaji2012mathematical, yang2019cr} across its key views. 
First, we construct a $2 \times 2$ affine transformation (i.e., rotation $+$ scaling) that transforms a triangle $f$ of the $i$-th part in the ``front'' view to the corresponding triangle in another view.
Second, we combine two methods: i) Kaji's local map~\cite{kaji2012mathematical} to compute a set of the triangle interpolated transformations $A_{f}$ and ii) Baxter's morphing technique~\cite{baxter2009n} to assemble them and optimize global transformations $B_{f}$, computed as
\vspace{2mm}
\begin{equation}
\argmin_{P} \sum_{f} \| B_{f} - A_{f} \|_{F}^{2} 
\label{eq:5}
\end{equation}
\begin{displaymath}
A_{f} = \sum_{j \in K} R_{f}^{j}(w_{j}) \cdot \exp{[w_{j} \log{S_{f}^{j}}(w_{j})]}
\end{displaymath}
\noindent
where $R_{f}^{j}(\alpha)$ and $S_{f}^{j}(\alpha)$ are the rotational matrix and scaling matrix, respectively, of the interpolated triangle shape, computed by the polar decomposition algorithm, at $\alpha \in \mathbb{R}$. 

This interpolation method appears to work well even if the silhouette of key views (e.g., left and right views) are not consistent, but any other interpolation technique, such as stroke-based interpolation~\cite{whited2010betweenit, yang2017context} or switching animation~\cite{willett2017triggering, willett2020pose}, could also be used with our framework.
\begin{figure}[t]
\centering
  \includegraphics[width=0.9\linewidth]{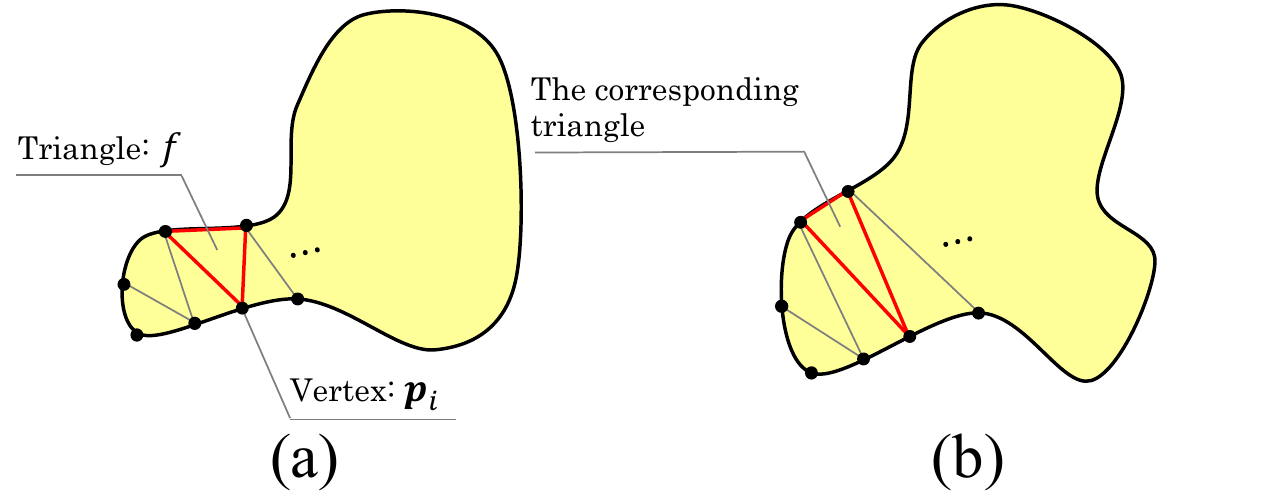}
  \caption{Triangle-based shape blending based on a set of 2D shapes from (a) front view and (b) another view.}
\label{fig:shape}
\end{figure}
%

%--------------------------------------------------------------------------
\subsubsection{Color-Space Interpolation}
In a similar way, we could simply blend color values of each part in key views on RGBA space $\bm{c}_{i}^{cur} \in (r, g, b, a)$ as follows:
\vspace{2mm}
\begin{equation}
\bm{c}_{cur}^{i} = \sum_{j \in K} w_{j} \bm{c}_{j}^{i} 
\label{eq:7}
\end{equation}
\noindent
where $\bm{c}_{i}^{j}$ is the $i$-th part's RGBA values in the $j$-th view.

%--------------------------------------------------------------------------
\subsubsection{Limited-Style Camera Control}
Our algorithm allows the user to freely look around the 2.5D model, but undesired popping artifacts may occur when the parts' Z-ordering changes as with previous 2.5D methods. Note that a certain amount of popping is actually expected in flip book cartoons~\cite{morimoto2019generating, willett2017triggering}. Then, we consider a method to mitigate an impression that popping effects give to users during rotation. One straightforward approach is to automatically generate cartoon styles from the user-specified camera rotation $R_{cur}$ in the view control panel. 
Existing studies on cartoon style~\cite{furukawa2017voice, kawamoto2008efficient, robert2019keyframe} modify input motion data such as motion-capture data by omitting some frames. While these methods can be suitable for reducing unintended popping, they run only on pre-recorded (offline) data. Therefore, we have considered two approaches in advance: (i)~a pre-recorded approach and (ii)~a real-time approach, and we concluded that the real-time method outperforms the offline method in terms of character design. 
In this paper, we compute XYZ-angles from an arbitrary viewpoint $R_{cur}$ and independently discrete them with a 10-degree interval. %Our system is simple but still useful for real-time use.
%~\cite{coutinho2016puppeteering, gois2015interactive, rivers20102}

%% file: sec/5_result.tex
\begin{figure}[t]
\centering
  \includegraphics[width=0.9\linewidth]{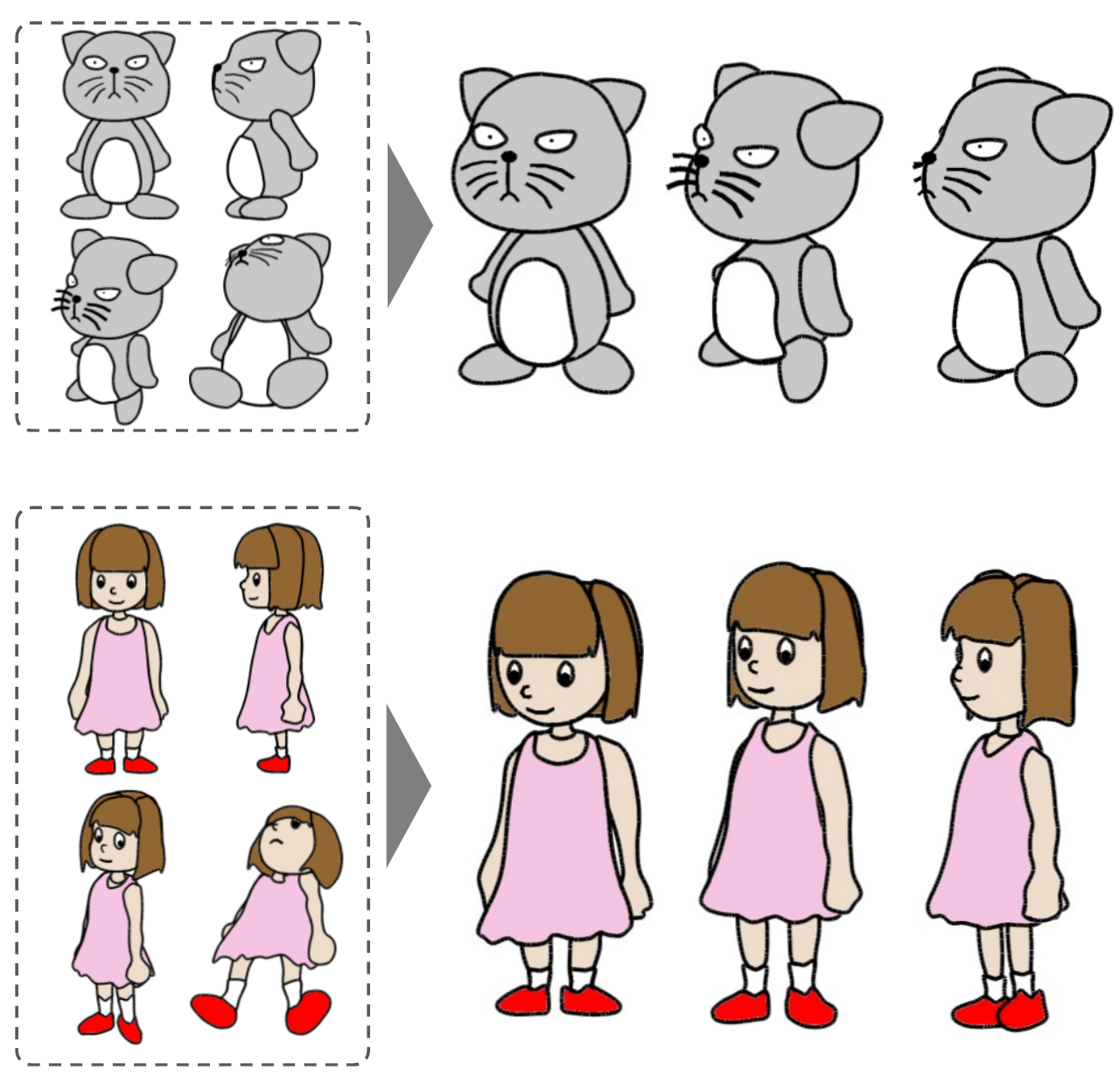}
  \caption{Examples of 2.5D graphics results. The input parts from several viewpoints were manually created by referring contour lines in cartoon character images (designed by artists using Harmony~\protect\shortcite{toonboom2020}).}
\label{fig:result}
\end{figure}
%

%--------------------------------------------------------------------------
\section{Results}
\label{sec:result}
Our prototype system was implemented on a $64$-bit 
Windows $10$ laptop (Intel\textcircled{\scriptsize R}Core$^{TM}$ i$7$-$5500$U CPU@$2.40$GHz and RAM $8.00$GB) using standard OpenGL and GLSL. The designed 2.5D cartoon models and their statistics are shown in Figure~\ref{fig:result} and Table~\ref{table:1}. As shown in these results, our system can make each 2.5D model within approximately $10$ mins (longer for the more complex models shown in Figure~\ref{fig:multilayer}).

%--------------------------------------------------------------------------
\begin{table}[t]
\centering
 \caption{Model Statistics.\vspace{-2mm}}
 \begin{tabular}{ll|c|c|c}
     \hline
     \raisebox{-0.3mm}{Model} && 
     \raisebox{-0.3mm}{\#\:Parts} & 
     \raisebox{-0.3mm}{\#\:Key Views} & 
     \raisebox{-0.3mm}{FPS} \\
     \hline
     \raisebox{-0.3mm}{Dog} &
     \raisebox{-0.3mm}{(Fig.~\protect\ref{fig:teasor})} & 
     \multicolumn{1}{r|}{\raisebox{-0.3mm}{$17$}} & 
     \multicolumn{1}{r|}{\raisebox{-0.3mm}{$8$\:}} & 
     \raisebox{-0.3mm}{$60+$}\\

     \raisebox{-0.3mm}{Cat} &
     \raisebox{-0.3mm}{(Fig.~\protect\ref{fig:result})} &
     \multicolumn{1}{r|}{\raisebox{-0.3mm}{$18$}} & 
     \multicolumn{1}{r|}{\raisebox{-0.3mm}{$4$\:}}  & 
     \raisebox{-0.3mm}{$60+$}\\
     
     \raisebox{-0.3mm}{Girl}  &
     \raisebox{-0.3mm}{(Fig.~\protect\ref{fig:result})} & \multicolumn{1}{r|}{\raisebox{-0.3mm}{$22$}} & \multicolumn{1}{r|}{\raisebox{-0.3mm}{$4$\:}}  & 
     \raisebox{-0.3mm}{$60+$}\\

     \raisebox{-0.3mm}{Panda} &
     \raisebox{-0.3mm}{(Fig.~\protect\ref{fig:3Dvs2.5D})}  & \multicolumn{1}{r|}{\raisebox{-0.3mm}{$9$}}  &
     \multicolumn{1}{r|}{\raisebox{-0.3mm}{$3$\:}}  & 
     \raisebox{-0.3mm}{$60+$}\\

     \raisebox{-0.3mm}{Bear}  &
     \raisebox{-0.3mm}{(Fig.~\protect\ref{fig:2Dvs2.5D})}  & \multicolumn{1}{r|}{\raisebox{-0.3mm}{$9$}}  & 
     \multicolumn{1}{r|}{\raisebox{-0.3mm}{$4$\:}}  & 
     \raisebox{-0.3mm}{$60+$}\\

     \raisebox{-0.3mm}{Boy} &
     \raisebox{-0.3mm}{(Fig.~\protect\ref{fig:failure})}   & \multicolumn{1}{r|}{\raisebox{-0.3mm}{$11$}}  &
     \multicolumn{1}{r|}{\raisebox{-0.3mm}{$2$\:}} & 
     \raisebox{-0.3mm}{$60+$}\\
     
     \raisebox{-0.3mm}{Ghost} &
     \raisebox{-0.3mm}{(Fig.~\protect\ref{fig:weight})}   & \multicolumn{1}{r|}{\raisebox{-0.3mm}{$6$}}  &
     \multicolumn{1}{r|}{\raisebox{-0.3mm}{$3$\:}} & 
     \raisebox{-0.3mm}{$60+$}\\ %\ref{fig:example}\&
     
     \raisebox{-0.3mm}{Walking Cycle} &
     \raisebox{-0.3mm}{(Fig.~\protect\ref{fig:animation})} & \multicolumn{1}{r|}{\raisebox{-0.3mm}{$11$}}  & \multicolumn{1}{r|}{\raisebox{-0.3mm}{$2 + 2$\:}}  & 
     \raisebox{-0.3mm}{$60+$}\\
     
     \raisebox{-0.3mm}{Raccoon Dog} &
     \raisebox{-0.3mm}{(Fig.~\protect\ref{fig:multilayer})}  & \multicolumn{1}{r|}{\raisebox{-0.3mm}{$26$}}  &
     \multicolumn{1}{r|}{\raisebox{-0.3mm}{$4$\:}}  & 
     \raisebox{-0.3mm}{$60+$}\\

     \raisebox{-0.3mm}{Alien} &
     \raisebox{-0.3mm}{(Fig.~\protect\ref{fig:multilayer})}  & \multicolumn{1}{r|}{\raisebox{-0.3mm}{$26$}}  &
     \multicolumn{1}{r|}{\raisebox{-0.3mm}{$4$\:}}  & 
     \raisebox{-0.3mm}{$60+$}\\

     \raisebox{-0.3mm}{Geracho} &
     \raisebox{-0.3mm}{(Fig.~\protect\ref{fig:multilayer})}  & \multicolumn{1}{r|}{\raisebox{-0.3mm}{$27$}}  &
     \multicolumn{1}{r|}{\raisebox{-0.3mm}{$3$\:}}  & 
     \raisebox{-0.3mm}{$60+$}\\
     \hline
 \end{tabular}
\label{table:1}
\end{table}
%\&\protect\ref{fig:weight}

%--------------------------------------------------------------------------
\subsection{Comparison of Rivers et al.} %Comparison
\label{sec:vsRivers}
We compared the results of our algorithm against Rivers et al.~\shortcite{rivers20102}. The existing 2.5D methods~\cite{coutinho2016puppeteering, kitamura2014modeling, gois2015interactive} compute anchor positions and shapes in the same way as Rivers et al., so this comparison is enough to show the effectiveness of our algorithm. 
%\red{Note that the existing 2.5D methods~\cite{coutinho2016puppeteering, kitamura2014modeling, gois2015interactive} compute anchor positions and shapes by the same mechanism as Rivers et al.} 

\subsubsection{About Anchor Interpolation} %Graphics
\begin{figure}[t]
\centering
  \includegraphics[width=0.9\linewidth]{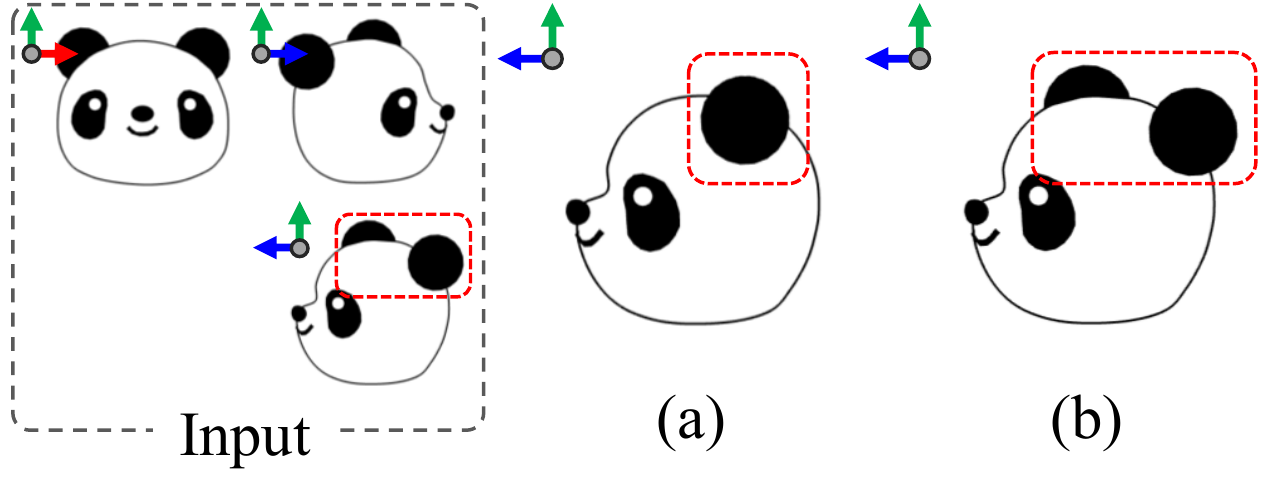}
  \caption{Comparing the anchor positions computed by (a)~the 3D-space interpolation (without 2D-space anchor distortion) and (b)~our 2.5D interpolation method (with 2D-space anchor distortion).}
\label{fig:3Dvs2.5D}
\end{figure}

\label{sec:vsAnchor}
First, we compared the results of our anchor estimation (with VDD) and 3D anchor results without the VDD mechanism, that is, $\bm{v}_{cur}^{i} = \Pi \: ( R_{cur} \bm{v}^{i})$. The results without VDD were mostly the same as Rivers's 3D anchor estimation. 
Figure~\ref{fig:3Dvs2.5D} shows examples of reproducing the input views using 2.5D models.
As expected, in case that the input parts (red square) do not correspond to any real 3D positions, the generated results without VDD do not match even the input and are difficult to handle the position displacements in its key views. On the other hand, by using our method, the plausible positions can still be obtained.
%interpolated results without VDD cannot represent the position displacements in its key views and are not suitable for cartoon models.
%2.5D cartoon models

%--------------------------------------------------------------------------

\begin{figure}[t]
\centering
  \includegraphics[width=0.9\linewidth]{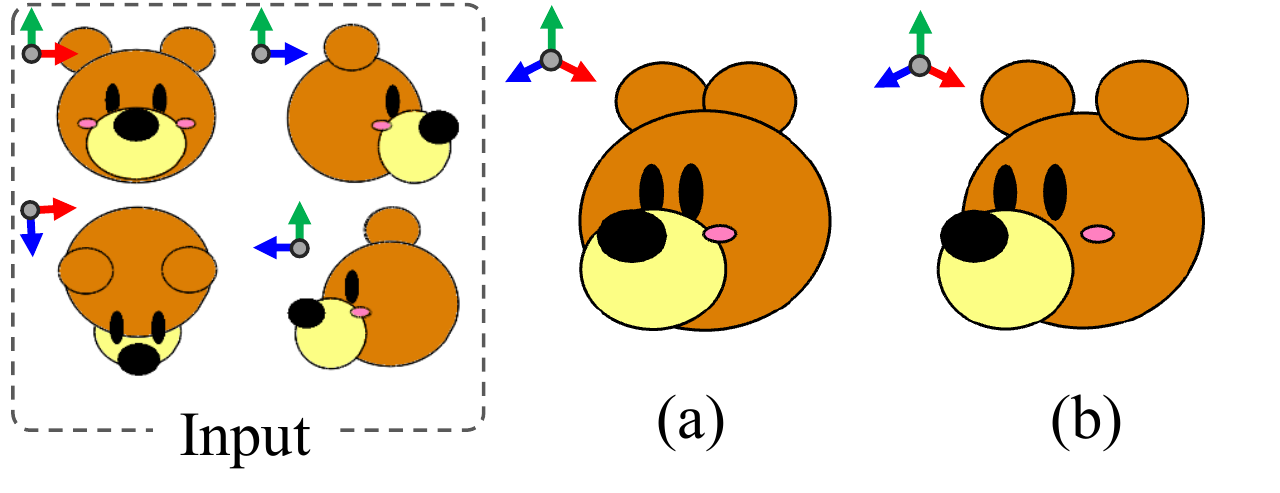}
  \caption{Comparing the anchor positions computed by (a)~the 2D-space interpolation (see Equation~(\protect\ref{eq:8})) and (b)~our 2.5D interpolation method.}
\label{fig:2Dvs2.5D}
\end{figure}
%--------------------------------------------------------------------------

Second, we compared the results of our anchor estimation against a simple 2D-space interpolation across its key views (without 3D-space anchor) as follows. This is a version that Rivers et al. used to compute anchor positions that do not correspond to any real 3D position, in any view (e.g., Mickey Mouse's ears).
%First, we compared the results of our anchor estimation against our own implementation of 2D-space interpolation (without 3D-space anchor). It is a 2D version of our anchor interpolation as follows:
%
%\vspace{2mm}
%\begin{equation}
%\bm{v}^{i}_{cur} = \hat{\bm{v}}_{i} + \sum_{j \in K} w_{j} (\bm{v}_{i}^{j} - \hat{\bm{v}}_{i})
%\label{eq:8}
%\end{equation}
%
\vspace{2mm}
\begin{equation}
\bm{v}^{i}_{cur} = \sum_{j \in k} w_{j} \bm{v}_{j}^{i}
\label{eq:8}
\end{equation}

\noindent
where $w_{j}$ is the weight value for the $j$-th views computed using Rivers's method. 
Figure~\ref{fig:2Dvs2.5D} shows the estimated results in oblique view. 
%The results are shown in Figure~\ref{fig:2Dvs2.5D}.
As expected, it is difficult for the 2D-space interpolation to represent 3D-like rotations across different views, such as out-of-plane rotation. 
In contrast, our anchor estimation achieves 3D-like movements because of the 3D-space anchor.

In summary, the method from Rivers et al. allows users to optionally select one of the two methods to compute the locations of all parts (including $Z$-ordering): (i)~using 3D-space anchors estimated from several key views, or (ii)~using 2D-space interpolations after removing 3D-associated anchors one by one, as in Equation~(\ref{eq:8}). 
% of all parts
%
When handling the view-specific anchor distortions (see Figure~\ref{fig:3Dvs2.5D}), the user has no choice but to use 2D-space interpolation (without the 3D anchor). However, this choice requires a lot of key viewpoints for making 3D-like movements (see Figure~\ref{fig:2Dvs2.5D}). In addition, their operations are complicated, making the process time-consuming and tedious. In contrast, our anchor estimation can easily deal with both the 3D-space movements and 2D-space distortions on all of the key views, so our system can produce significantly better results from a small number of key viewpoints.
%and is not suitable 

\begin{figure}[t]
\centering
  \includegraphics[width=0.88\linewidth]{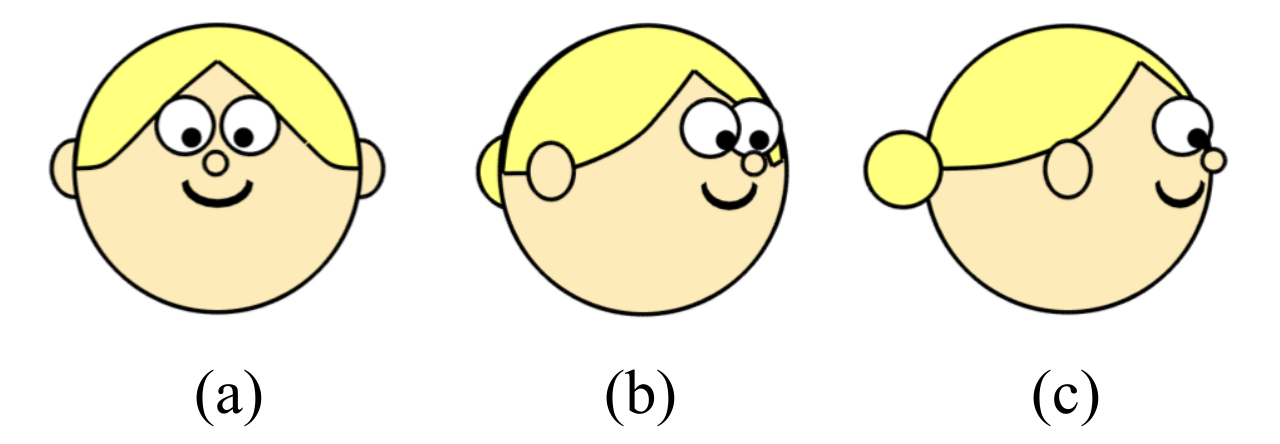}
  \caption{An example of (b)~2.5D graphics from (a, c)~two key views. The input model was manually designed while referring to \cite{rivers20102}.}
  %\caption{An example of failure 2.5D graphics (a parting position in hair) from (a, c)~two key-views. The reason is that the parts' shapes are determined by a 2D-space interpolation method.}
\label{fig:failure}
\end{figure}

%--------------------------------------------------------------------------
\subsubsection{About Shape Interpolation}
\label{sec:vsShape}
In determining all of the parts' shapes in a novel view, Rivers et al. select $k$-nearest key views on the 2D parameterized angle space and interpolate the key shapes that consist of silhouette lines by using a linear vertex interpolation. However, their method cannot prevent local distortions in the intermediate shapes because the intermediate shapes do not consider the rotational and stretching components. 
In contrast, our interface enables users to input (and edit) triangulated models, so it is possible to minimize the distortion of the intermediate shapes (see Figure~\ref{fig:failure}).
However, our system, like Rivers's method, employs 2D-space interpolation techniques. 
It is still difficult to translate all the vertices of each part along 3D-like paths. It might be better to interpolate parts' shapes by adding 3D-space information, if necessary. 
Second, as mentioned in Section~\ref{sec:management}, our interface allows users to manually deform each part's shape (i.e., input data) and build a vertex correspondence between key views. This manual function is simple but effective for improving the visual quality of 2.5D graphics. In addition, we investigated existing methods to fully automatically build a silhouette correspondence between multi-view 2D shapes~\cite{sederberg1992physically, baxter2009compatible}, but we refrain from involving them in our interface for the time being. This is because most of them do not support highly concave shapes (e.g., a parting position of the hair) and regions of partial occlusion. We will plan to consider semi-automatic methods to build correspondence between 2D shapes specialized for our problem.
% at an intermediate view

%--------------------------------------------------------------------------
\begin{figure}[t]
\centering
  \includegraphics[width=0.85\linewidth]{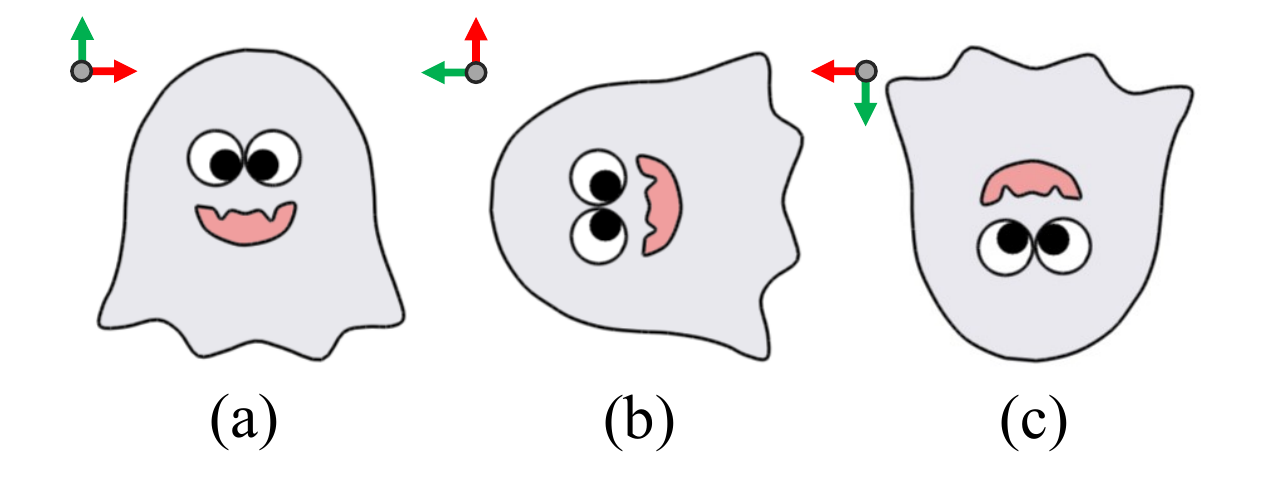}
  \caption{An experimental set of key views that consist of (a)~the front view (regular attitude) and two attitudes rotated at (b)~$90^\circ$ and (c)~$180^\circ$ around the $Z$-axis without changing the position from the front view (i.e., roll rotation).}
\label{fig:example}
\end{figure}

\begin{figure}[t]
%\centering
\hspace*{-5mm}
  \includegraphics[width=1.08\linewidth]{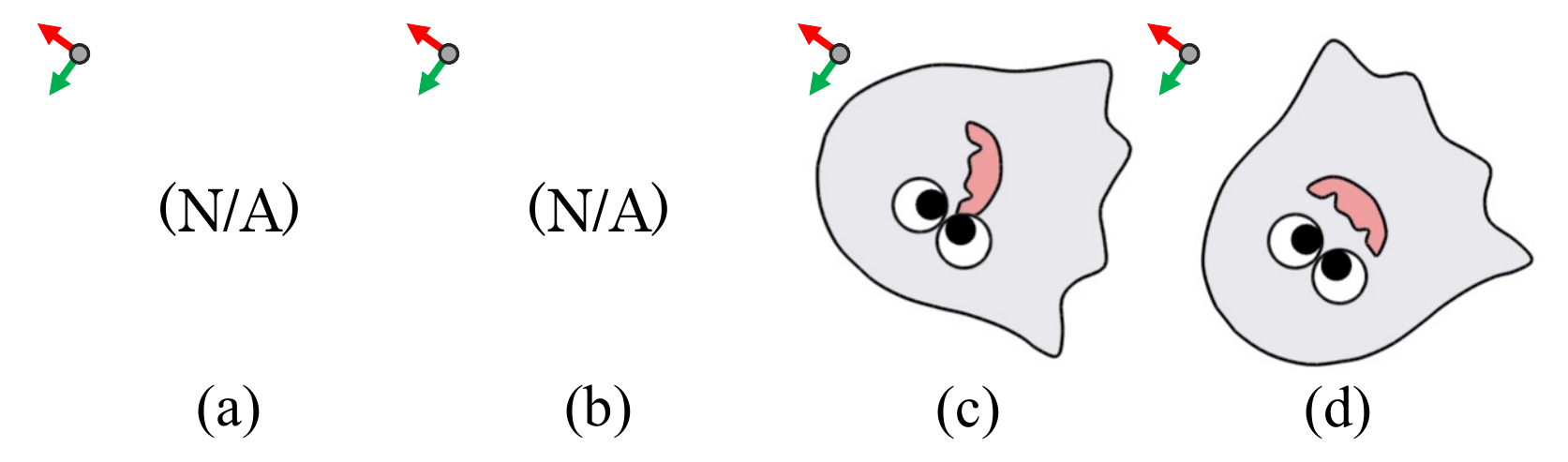}%0.785
  \caption{Comparing the weight computation results of three methods; (a)~Rivers et al.~\shortcite{rivers20102}, (b)~Rademacher~\protect\shortcite{rademacher1999view}, (c)~Koyama et al.~\protect\shortcite{koyama2013view}, and (d)~our method, defined by a $135^\circ$ rotation about the $Z$-axis.}
\label{fig:weight}
\end{figure}

%--------------------------------------------------------------------------
\begin{figure*}[t]
\centering
  \includegraphics[width=0.9\linewidth]{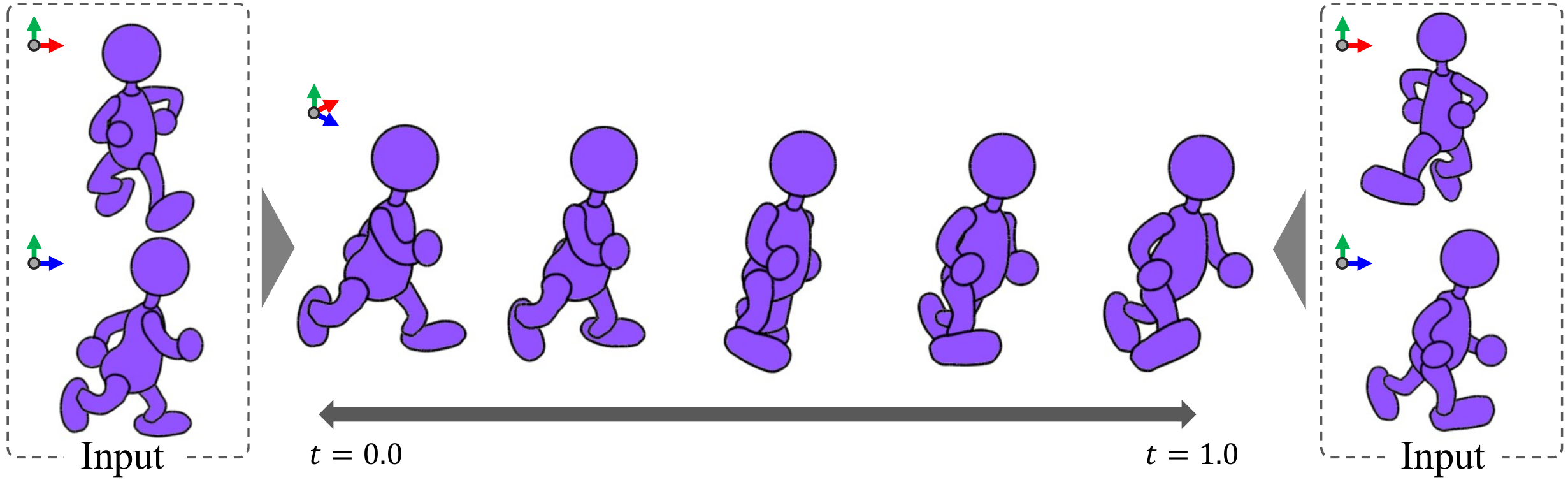}
\caption{An example of animation results between two 2.5D cartoon models (from the frontal view and the lateral view) at time $t \in \{0.0, 1.0\}$. The input models are created by referring to a walking cycle from \cite{rademacher1999view}.}
\label{fig:animation}
\end{figure*}

%--------------------------------------------------------------------------
\subsection{Comparison of Weight Computation}
\label{sec:weight}
We applied both the proposed weight computation and other methods to a 2.5D cartoon model generated by our interface. 
Figure~\ref{fig:weight} shows blending results generated by (a)~Rivers's method~\shortcite{rivers20102}, (b)~Rademacher's method~\shortcite{rademacher1999view}, (c)~Koyama's method~\shortcite{koyama2013view}, and (d)~our method. Note that the references (character poses) consist of the front view (regular attitude) and two attitudes rotated at $90^\circ$ and $180^\circ$ around the $Z$-axis without changing the view position from the front view (i.e., roll rotation), as shown in Figure~\ref{fig:example}. 
In order to compute blending weights, Rivers's system constructs a Delaunay triangulation of key viewpoints in a 2D parameterization of the angle space (i.e., yaw and pitch), and then finds the $k$-nearest key views on the space. 
Similarly, Rademacher's method directly constructs a convex hull of the key viewpoints in 3D space. 
Therefore, in the case of the above key viewpoints which have the same yaw and pitch angles and positions, these methods cannot compute the blending weights at all (see Figure~\ref{fig:weight}(a, b)). 
Koyama's method relies only on an angle between the viewing ray and key viewpoint position, so their system equally blends all of the above data and shows an inappropriate rotation result, as shown in Figure~\ref{fig:weight}(c).
In summary, the previous methods cannot consider differences of camera rotation; hence, the position information of viewpoint is unsuitable for computing the blending weights. 
On the other hand, by using our blending method, plausible results can still be obtained (see Figure~\ref{fig:weight}(d)).

%--------------------------------------------------------------------------
\subsection{2.5D Cartoon Animation Over Time}
For generating cartoon animations from an arbitrary viewpoint, one straightforward approach is to prepare 2.5D models for every frame, but this requires a lot of time. Hence, as with traditional 3D animation systems, our 2.5D graphics can easily be extended to create an animation from a small number of the 2.5D models at selected frames by linearly interpolating all components of 2.5D models (i.e., 3D anchor positions, the distortions, and the shapes), as shown in Figure~\ref{fig:animation}. Of course, this system can further improve the quality by combining with existing systems, such as a skeletal deformer~\cite{hornung2007character, coutinho2016puppeteering}, in the 3D anchor interpolation process.

%We have used our interface to generate models from a variety of found 2D images. These are shown in Figures 20 and 21 and in the accompanying video. A typical modeling session lasted less than ten minutes (longer for the more complex models shown in Figure 21). Results created by user testers are presented in Sec- tion 7. Source material included drawings from a children’s book, user-created 2D drawings, concept artwork, and cartoons.

%% file: sec/6_userstudy.tex
\section{User Study}
\label{sec:userStudy}
We conducted a user study to gather feedback on the quality of the 2.5D cartoon models and our modeling tool (e.g., advantages and limitations) from participants. This is because the previous methods~\cite{coutinho2016puppeteering, gois2015interactive, kitamura2014modeling, rivers20102} focused only on cartoon-like representation techniques, and subjective impressions have never been studied.
\subsection{Procedure}
\label{sec:procedure}
We invited 10 participants (P1, P2, $\cdots$, P10) aged 20--40 years ($Avg.=25.3$, $Var.=3.13$) to evaluate the usability of our system, using a standard mouse as an input device. 
%Note that we determined the number of participants (sample size) based on the professional standards for this type of study within the human-computer interaction community~\cite{caine2016local}.
Each participant was asked to fill out a form asking about their experience with designing 2D/3D models using commercial software. 
%
%++++++++++++++++++++++++++++++++++
%both
%hori++++++++++++++++++++++++++++++++++
P1 had extensive experience in creating 3D models with Side FX Houdini and Autodesk Fusion $360$ ($> 8$ years), and drawing cartoon tools with Clip Studio Paint ($> 3$ years) to create video games as a hobby.
%
%toby & asai & noh+++++++++++++++++++++
P2–4 had professional programming experience, building applications for commercial and research purposes. They also had prior experience of 3D modeling software, such as RealityCapture and Blender ($> 3$ years) and image processing software, such as Adobe Photoshop ($> 2$ years).
%
%3D modeling only%%%%%%%%%%%%%%%%%%%%%%
%hashimoto & takahashi & Nakashima+++++
P5–7 were experienced users of 3D modeling software, such as Pixologic ZBrush and Blender ($> 2$ years) but had no prior experience in drawing cartoons.
%
%suetake & chou++++++++++++++++++++++++
P8–9 had moderate amounts of experience drawing cartoons with Paint Tool SAI or Adobe Illustrator ($> 1$ year). 
%
%neather%%%%%%%%%%%%%%%%%%%%%%%%%%%%%%%
%kubota++++++++++++++++++++++++++++++++
P10 had basic knowledge of the JavaScript and C programming languages that are used in code editor, but no graphical design experience.

%++++++++++++++++++++++++++++++++++
First, we gave them a brief overview of our modeling tools. The instructor explained a step-by-step tutorial to familiarize the participants with this modeling framework. After an overview explanation, they could smoothly design multi-view 2D parts (i.e., location and shape) using our system. 
Next, we also provided them with a 2D layered model in the front view and asked them to keep designing their own 2.5D cartoon models until they were satisfied. 
At the end of the 2.5D cartoon model creation, the participants filled out a questionnaire consisting of four questions about our system's usability (see Table~\ref{table:2}:\:Question Items) using a seven-point Likert scale (from $1$:\textit{\:Extremely dissatisfied} to $7$:\textit{\:Extremely satisfied}). The purpose of these questions was to analyze their subjective impressions.

%--------------------------------------------------------------------------
\subsection{Observations and User Feedback}
\label{sec:observation}
Table~\ref{table:2} shows the post-experiment questionnaire results, giving the mean values ($> 4$) and standard deviations. The participants' comments regarding the proposed 2.5D cartoon model are summarized below. 
% and their previous experience with anime creation

\noindent
\begin{itemize}
%\setlength{\leftskip}{-3mm}
%操作、パーツを並べるだけで簡単(hori & chou & toby)++++++++++++++++++++
\item P1\&10: \textit{I thought that this design logic is clear and felt relaxed to learn it.}
\item P2: \textit{The user interaction was similar to that of the modeling software that I am used to (e.g., Cartoon Animator 4~\shortcite{reallusion2020cartoon} and Live 2D~\shortcite{live2019}). Using different operations is also intuitive. The user interface provides only relevant information and is easy to grasp.}
%アニメーションに便利（Toby & takahashi）++++++++++++++++++++++++++++++
\item P2\&P7: \textit{It looks very convincing and is very easy to use, and I think that this software is quite useful for making short animations (e.g., flash animation).}
%タイムラグ（nakashima & hashimoto）+++++++++++++++++++++++++++++++++++
\item P5\&P6: \textit{I like how fast it is to generate 2.5D models when clicking the Calc button, so I can concentrate on the 2.5D model modeling without getting stressful.}
%絵コンテ作業（sue）+++++++++++++++++++++++++++++++++++++++++++++++++++
\item P8: \textit{I would like everybody, especially animators, to use 2.5D cartoon models from character concept art images when making their own storyboard.}
%CADっぽくて想像しやすい(sue, chou)++++++++++++++++++++++++++++++++++++
\item P8\&9: \textit{I have no experience in 3D-model building, but if I imagine the process, I feel it tedious to have every part correctly located. In contrast, to edit 2D component from different views like CAD may be easier for me.}
\end{itemize}

Overall, the participants reported that the process of 2.5D character modeling (i.e., 2D-space operation) is straightforward to use and useful. A possible reason is that the proposed 2.5D modeling tool enables users to easily handle 3D-like rotations without 3D-space modeling. We think that our 2.5D modeling system can be used as a ``base'' tool for cartoon design, and it might be interesting to explore the possibility of incorporating other functions into our tool. There were some requests from participants to add functions, as follows:
% for creating cartoon animation

\begin{table}[t]
%\centering
\hspace*{-2mm}
 \caption{Results of the post-experiment questionnaire.}
 \begin{tabular}{c|lcc}
     \hline
     \raisebox{-0.3mm}{\#} &
     \raisebox{-0.3mm}{Question Items} & 
     \raisebox{-0.3mm}{Mean} & 
     \raisebox{-0.3mm}{SD} \\
     \hline
     \raisebox{-0.3mm}{1}  &  
     \raisebox{-0.3mm}{It was easy to learn how to use.} & 
     \raisebox{-0.3mm}{$6.20$} & 
     \raisebox{-0.3mm}{$0.60$}\\
     \raisebox{-0.3mm}{2} &  
     \raisebox{-0.3mm}{It was comfortable to use.}  & \raisebox{-0.3mm}{$6.20$} & 
     \raisebox{-0.3mm}{$0.75$}\\
     \raisebox{-0.3mm}{3} &  
     \raisebox{-0.3mm}{I didn't feel stress after the design task.} & 
     \raisebox{-0.3mm}{$5.90$} & 
     \raisebox{-0.3mm}{$0.83$}\\
     \raisebox{-0.3mm}{4} &  
     \raisebox{-0.3mm}{The quality of 2.5D models was satisfactory.} & 
     \raisebox{-0.3mm}{$6.10$} & 
     \raisebox{-0.3mm}{$0.70$}\\
     \hline
 \end{tabular}
\label{table:2}
\end{table}
%I satisfied the quality of $2.5$D graphics.
%     \raisebox{-0.3mm}{The quality of 2.5D graphics was satisfactory.} & 

\noindent
\begin{itemize}
%\setlength{\leftskip}{-3mm}
%カメラの操作をlimitedにしたい(hori)++++++++++++++++++++++++++++++++++++
%\item P1: \textit{This system allows the user to freely look around the 2.5D model, but I further want to convert original camera transitions into classic cartoons-style (at $24$ fps) by omitting some frames.}
%テキスト機能（asai）+++++++++++++++++++++++++++++++++++++++++++++++++++
\item P3: \textit{This software does not show much textual information, but adding them might make it easy to aid 2.5D character modeling.}
%位置の補正機能（sue, hashi）+++++++++++++++++++++++++++++++++++++++++++
\item P6\&8: \textit{When roughly placing 2D components on the modeling panel, I want the system to automatically identify groups and align individual components one by one, for example vertical and horizontal alignment~\cite{Xu2015gaca}.}
\end{itemize}

According to these comments, the participants also identified several issues with the current implementation, but we found those not to be serious problems, and it is possible to further improve user experience with an engineering effort.

%--------------------------------------------------------------------------
\section{Professional Feedback}
\label{sec:professional}
We also asked two professional animators working in cartoon production about the quality and the application possibility of 2.5D cartoon models designed by our prototype system. Their comments are summarized below.

\noindent
\begin{itemize}
\item \textit{This 2.5D cartoon modeling method has the potential to drastically improve the efficiency of classical cartoon design in future.}
\item \textit{The current system's operations are very intuitive, so I thought that 2D artists (without 3D modeling skill) can easily make their own 2.5D models with no problems.}
\item \textit{The position correction (anchor interpolation) and the 2D shape interpolation are simple but still useful for animators to generate a choice of classical cartoon's representation such as hybrid animations.}
\end{itemize}

In their experience, 3D modeling systems can be difficult to use, even for experienced artists, and do not retain the detail of 2D drawings (e.g., Mickey Mouse's ear) at all. Thus, the 3D model usage is limited in classical cartoon productions (e.g., background and effects design). In contrast, we confirm that 2.5D cartoon models and our tool are expected to efficiently design cartoon animations. They also commented that the 2.5D graphics could be very helpful to their manual drawing tasks.
%In summary, 

\noindent
\begin{itemize}
%\setlength{\leftskip}{-3mm}
%\item \textit{The 2.5D cartoon models can also be useful for supporting animators' drawings (e.g., in-betweening) by tracing the screen of the generated 2.5D models on sheets of paper.}
\item \textit{With the proposed system, animators can easily generate various character images composed at an oblique angle. I think that 2.5D models can support animators' manual drawings by tracing the screen of the generated 2.5D models on sheets of paper.}
\item \textit{The 2.5D models can also be used for doing proper in-betweening. As with the limited-camera control, I expect to incorporate an automatic function to select an optimal set of frames from interpolated animated results (limited-cartoon style keyframing)~\cite{kawamoto2008efficient, robert2019keyframe}.}
\end{itemize}
%he 2.5D models enable to make an animation from a small number of the 2.5D models, so t

In summary, the professional animators also suggested some example usage scenario from their point of view. As far as we know, there is not enough computer-assisted software to design classical cartoon characters, so we will take this opportunity to spread the 2.5D graphics systems into the animator field in the future.
%It may be interesting to extend

%As far as we know, there is not enough computer-assisted software to design classical cartoon characters, so we should further extend and spread it into the animator field in the future. 

%まとめると，プロのアーティストの目線からも，本システムを用いることで2.5D モデルの設計が容易であることや，2.5D モデルの可能性が期待されていることがわかった．
%また，セルアニメ作品向けのサポートが少ない現状からしても，本インターフェースのようなシステムはセルアニメ作品の新たな制作方法を提示できるのではないかと考えられる．

\begin{figure}[t]
\centering
  \includegraphics[width=0.95\linewidth]{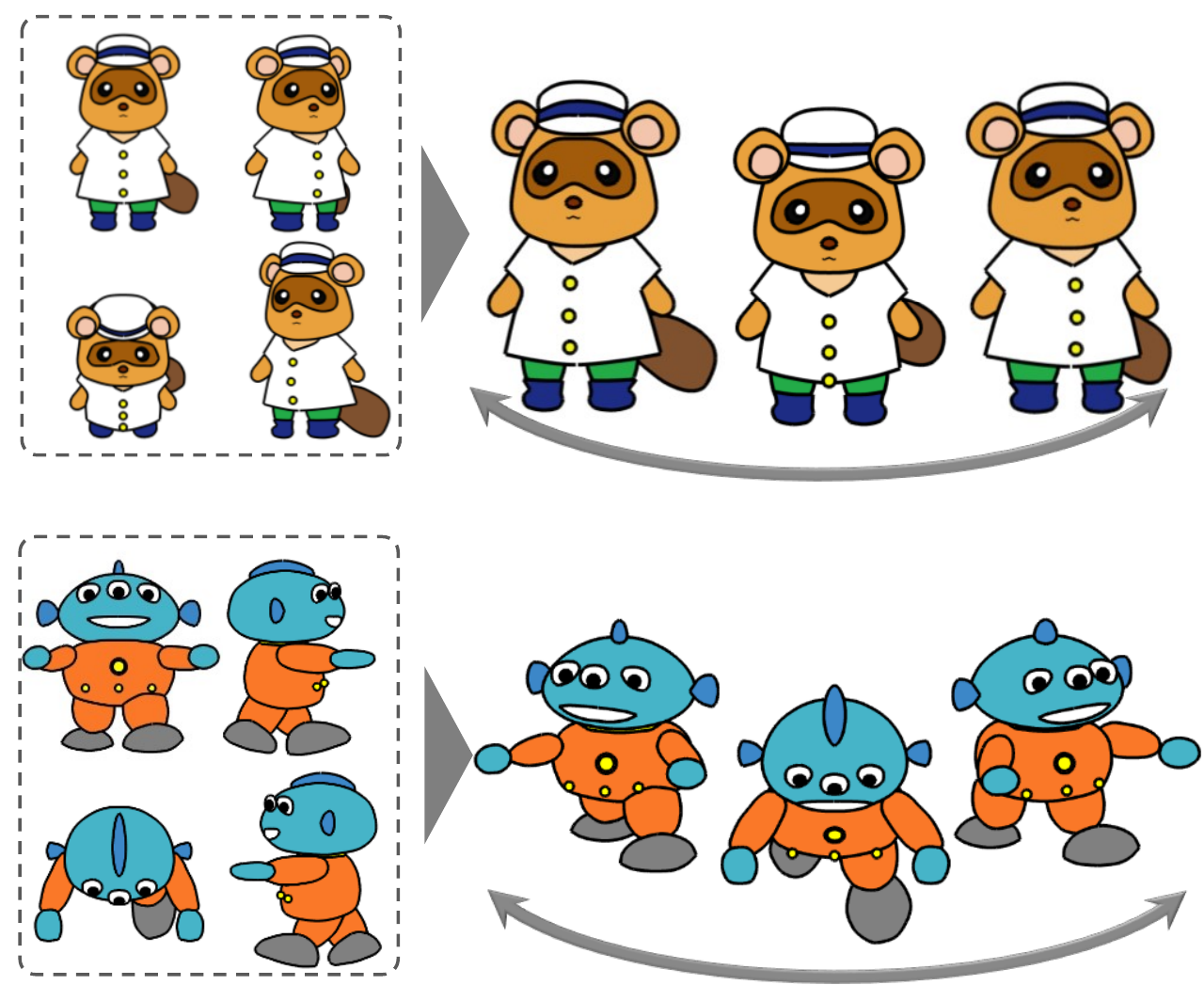}
  \caption{More complex models created using our interface. From top to bottom, the input models are designed by reference to Raccoon Dog from \cite{kitamura2014modeling}; Alien from \cite{rivers20102}.}

%  \includegraphics[width=0.95\linewidth]{fig/complexModel.pdf}
%  \caption{More complex models created using our interface. From top to bottom, the input models are designed by reference to Raccoon Dog from \cite{kitamura2014modeling}; Alien from \cite{rivers20102}; Geracho from Cocotama (\url{https://toy.bandai.co.jp/series/cocotama/}).}

\label{fig:multilayer}
\end{figure}
%\copyright \copyright OLM digital Inc.

%% file: sec/7_limitation.tex
\section{Limitations and Future Work}
\label{sec:limitation}
Although our limited-style camera control can mitigate a bit of an impression that popping artifacts are given to users during rotation, this is not a solution to the root of Rivers's popping problem. To address this issue, we plan to support partial occlusions of some shapes (e.g., concave regions) in future. For example, it might be better to allow users to (1)~make several anchor points inside each part, (2)~decompose one part into many sub-parts, or (3)~manually change a sprite switching time, if necessary. 
%the root of the 
In addition, the current appearance is computed only by RGBD color blending and the users are not allowed to apply shading and texture mapping methods. Thus, a potential future work is to incorporate better shading method~\cite{gois2015interactive} and texture mapping method~\cite{debevec1998efficient}.

While striving for a simple user interface with minimal user input, our system might not accommodate more complicated operations (e.g., using a special device such as a pen tablet). Achieving a balance between simplicity and functionality might be a good research topic.

%Our issue with our current system is that the appearance is computed by RGBD color blending and the users are not allowed to apply shading and texture mapping methods. It might be interesting to explore the possibility of extending our system to include other processes, such as shading~\cite{gois2015interactive} and texture mapping~\cite{debevec1998efficient}, in the future.

%% file: sec/8_conclusions.tex
\section{Conclusion}
\label{sec:conclusion}
This paper has presented a method to design 2.5D cartoon models from multi-view 2D shapes as follow: We have newly formulated the concept of Rivers's method~\shortcite{rivers20102} by combining it with VDD techniques~\cite{chaudhuri2004system, chaudhuri2007reusing, koyama2013view, rademacher1999view}. 
%This paper has presented an interactive method to design 2.5D cartoon models from multi-view 2D shapes as follow: 
Based on the proposed 2.5D graphics, we can easily and quickly emulate a 3D-like movement while preserving the view-space effects in a 2D cartoon. Hence, we believe that our formulation will be a new step toward the acceleration of research in classical 2D cartoon animations in the future.

\section*{Acknowledgement}
%Removed for review.

We would like to thank Yu Takahashi (OLM, Inc.) and Tatsuo Yotsukura (OLM Digital, Inc.) for their valuable comments and helpful suggestions. The characters in Figure~\ref{fig:teasor}, \ref{fig:screenshot}, and \ref{fig:result} are copyrighted by OLM.

%The authors would like to thank Dr. Yuhua Li for providing the MATLAB code of the \textit{BEPS} method.
%The authors would also like to thank the anonymous referees for their valuable comments and helpful suggestions. The work is supported by the \grantsponsor{GS501100001809}{National Natural Science Foundation of China}{https://doi.org/10.13039/501100001809} under Grant No.: ~\grantnum{GS501100001809}{61273304}
%21 and ~\grantnum[http://www.nnsf.cn/youngscientists]{GS501100001809}{Young Scientists' Support Program}.
%\end{acks}